\documentclass[prl,twocolumn,a4paper,amsmath,floatfix]{revtex4}
\usepackage[utf8]{inputenc}
\usepackage{graphicx}
\usepackage{amsmath}

\usepackage{color}
\newcommand{\HexL}{Hex\textsubscript{L}}
\newcommand{\HexS}{Hex\textsubscript{S}}
\newcommand{\HexLplus}{Hex\textsubscript{L}\textsuperscript{+}}

\begin{document}

\title{Self-assembly of dodecagonal and octagonal quasicrystals in hard spheres on a plane}
\author{Etienne Fayen$^1$, Marianne Imp\'eror-Clerc$^1$, Laura Filion$^2$, Giuseppe Foffi $^1$, Frank Smallenburg$^1$}
\affiliation{$^1$Universit\'e Paris-Saclay, CNRS, Laboratoire de Physique des Solides, 91405 Orsay, France\\
$^2$Soft Condensed Matter, Debye Institute of Nanomaterials Science, Utrecht University, Utrecht, Netherlands \\}
\begin{abstract}
    Quasicrystals are fascinating structures, characterized by strong positional order but lacking the periodicity of a crystal. In colloidal systems, quasicrystals are typically predicted for particles with complex or highly specific interactions, which makes experimental realization difficult. Here, we propose an ideal colloidal model system for quasicrystal formation: binary mixtures of hard spheres sedimented onto a flat substrate. Using computer simulations, we explore both the close-packing and spontaneous self-assembly of these systems over a wide range of size ratios and compositions. Surprisingly, we find that this simple, effectively two-dimensional model systems forms not only a variety of crystal phases, but also two quasicrystal phases: one dodecagonal and one octagonal. Intriguingly, the octagonal quasicrystal consists of three different tiles, whose relative concentrations can be continuously tuned via the composition of the binary mixture. Both phases form reliably and rapidly over a significant part of parameter space, making hard spheres on a plane an ideal model system for exploring quasicrystal self-assembly on the colloidal scale.
\end{abstract}

\maketitle

\section{Introduction}

Hard spheres are arguably the most fundamental model system in colloid science. 
The colloidal equivalent of marbles, hard spheres only interact when colliding, but despite this simplicity exhibit nearly all important aspects of phase behavior. As such, colloidal hard spheres have been instrumental in enhancing our understanding of crystal nucleation \cite{o2003crystal,gasser2001real}, crystallization in confinement \cite{de2015entropy, wang2021binary, fortini2006phase, curk2012layering, jung2020confinement, fu2017assembly, duran2009ordering}, two-dimensional melting \cite{thorneywork2017two, bernard2011two}, glassy dynamics \cite{foffi2004alpha, weeks2000three, zaccarelli2009crystallization, marin2020tetrahedrality, boattini2020autonomously}, crystal defects \cite{pronk2001point,van2020high,van2017diffusion},  among many others.  Their important role in soft matter science stems not only from their theoretical simplicity  and the ease and speed at which they can be simulated \cite{smallenburg2022efficient,klement2019efficient},  but also from the fact that they can be quantitatively explored in the lab \cite{pusey1986phase, yethiraj2003colloidal, dullens2006colloidal, royall2013search}.  

One aspect of colloidal phase behavior where hard sphere have thus far not proven suitable as a model system is the formation of quasicrystals. These exotic structures, characterized by a forbidden symmetry incompatible with normal crystalline order, have been predicted or observed to form in a variety of soft-matter systems consisting of nanoparticles or macromolecules \cite{dotera2011quasicrystals, talapin2009quasicrystalline, gillard2016dodecagonal, duan2018stability}, but have so far remained elusive in colloidal particles on the micrometer scale. 
This is unfortunate, as a colloidal model system that reliably forms quasicrystals would be ideal for the real-time study of quasicrystal self-assembly. 
In computer simulations of colloidal soft matter, quasicrystals are typically found in systems with highly specific interactions -- such as oscillatory potentials, patchy interactions, and square-shoulder repulsion \cite{dotera2011quasicrystals, van2012formation, zu2017forming, damasceno2017non, gemeinhardt2019stabilizing, malescio2021self} -- which are hard to reproduce in the lab. While complex quasicrystal approximants have been found to self-assemble in simulations of polydisperse mixtures of hard spheres \cite{bommineni2019complex}, and finite clusters with icosahedral symmetry have been shown to form in spherical confinement, thus far hard-sphere systems have not been found to be capable of forming a quasicrystal.

Here, we propose a possible solution to this lack of a colloidal test case for quasicrystal formation by examining the self-assembly of binary mixtures of hard spheres sedimented onto a flat plane. We find that this simple quasi-two-dimensional system exhibits an amazingly rich self-assembly behavior, forming not only six periodic crystal phases, but two quasicrystals as well:  one dodecagonal and one octagonal. Although dodecagonal quasicrystals are relatively common in soft matter models \cite{van2012formation, dotera2014mosaic, pattabhiraman2015stability, talapin2009quasicrystalline, ye2017quasicrystalline, zu2017forming}, octagonal ones are much more rare \cite{zu2017forming, damasceno2017non, malescio2021self, gemeinhardt2019stabilizing}. Moreover, unlike previously observed eight-fold quasicrystals made up of two tiles, the octagonal tiling we observe here is composed of three distinct tiles, whose relative composition can be directly tuned by changing the fraction of small spheres in the system. Both quasicrystal phases form reliably and rapidly over a significant part of parameter space. 
As binary hard spheres on a plane are directly experimentally realizable, to the point where they quantitatively match simulations \cite{thorneywork2014communication, thorneywork2017two}, this discovery identifies  an ideal model system for studying essentially all properties of quasicrystals, including their structure,  nucleation, melting, and defects dynamics.

\begin{figure}[b]
    \centering
    \includegraphics[width =\linewidth]{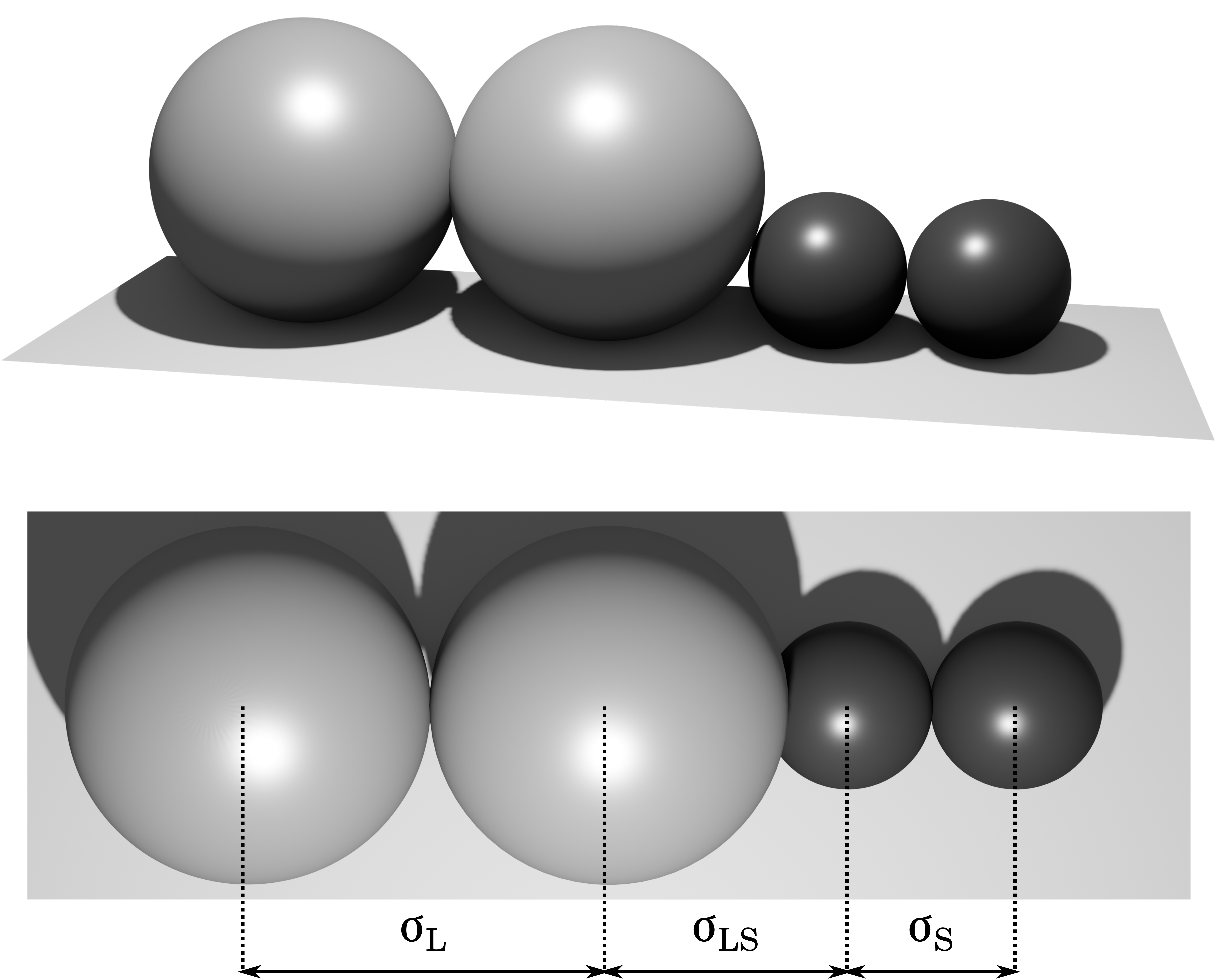}
    \caption{Schematic depiction of the model. 3D hard spheres lying on a flat surface (top) can be interpreted as an equivalent 2D system of non-additive hard disks (bottom). Disks of the same type behave like standard hard disks, while the closest distance between disks of different types $\sigma_{LS}$ is smaller than the sum of the radii.}
    \label{fig:spheres_on_plane}
\end{figure}

\section{Model}

We consider mixtures of hard spheres of two different sizes constrained to lie on a flat plane. As the particles are contained to move in only two dimensions, particles of equal size interact simply as hard disks.  However, as illustrated in Fig.  \ref{fig:spheres_on_plane}, for spheres of different sizes, a small amount of overlap of the 2D projection of the particles is allowed.  Specifically, when viewed from the top the distance of closest approach between a large particle of diameter $\sigma_L$ and a small particle of size $\sigma_S$ is  given by the geometric mean of their diameters: $\sigma_{LS} = \sqrt{\sigma_L \sigma_S}$. 

The phase behavior of a mixture of $N_L$ large spheres and $N_S$ small spheres confined to a substrate with area $A$ is controlled by three parameters: the size ratio $q = \sigma_S / \sigma_L$, the fraction of small disks $x_S = N_S / (N_L+N_S)$, and the area fraction $\eta = (N_S \sigma_S^2 + N_L \sigma_L^2)\pi/4A$. Note that since some overlap is allowed between different species, the total area fraction may exceed 1 in some cases.

\section{Infinite pressure}

\begin{figure*}
\raggedright
    \includegraphics[width=\textwidth]{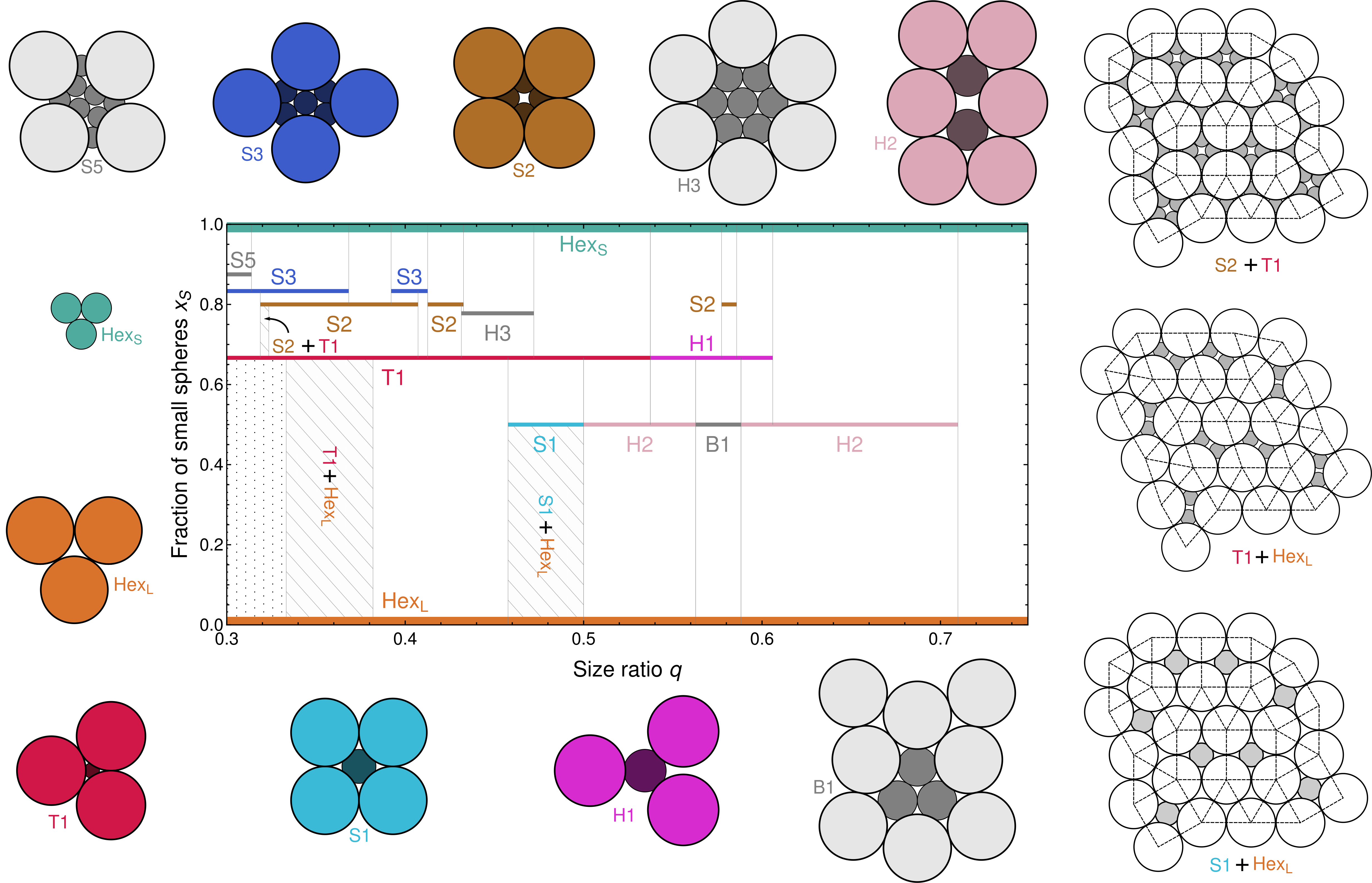}
    \caption{Infinite-pressure phase behavior of binary mixtures of spheres on a flat plane, as a function of the size ratio $q$ and fraction of small particles $x_S$. Phases are labeled following the naming scheme of Refs. \cite{fayen2020infinite, likos1993complex}. The white regions correspond to coexistence regions between the phases directly above and below. The hashed and dotted areas indicate regions where these two phases can form random tilings or a lattice gas, respectively. 
    Examples of finite patches of the three possible random tilings, corresponding to the hashed regions in the diagram, are displayed on the right.}
    \label{fig:infinitepressure}
\end{figure*}

%To obtain a first impression of the crystals we might expect to find in these mixtures, we first examine their close-packing behavior in the limit of infinite pressure. In previous work \cite{fayen2020infinite}, we determined the best-packed structures for non-additive binary mixtures of hard disks over a range of size rations and compositions, by constructing a library of candidate crystal structures and finding -- for each combination of $q$ and $x_S$ -- the best-packed phase or coexistence of phases. This then corresponds to the expected phase behavior at infinite pressure. 

Even for simple binary mixture in 2D, the number of different ordered structures that can emerge can be quite large and difficult to enumerate. To obtain an impression of the crystals we might expect to find, we used a technique specifically designed to the detect the close-packing crystal structures that would form in the limit of infinite pressure. To this end, we followed Ref.~\onlinecite{fayen2020infinite}, where we determined the best-packed structures for non-additive binary mixtures of hard disks over a range of size ratios and compositions, by constructing a library of candidate crystal structures and finding -- for each combination of $q$ and $x_S$ -- the best-packed phase or coexistence of phases.

Here, we follow the same approach to map out the infinite pressure phase diagram of spheres on a plane. Due to the two-dimensional confinement, our systems are equivalent to non-additive binary mixtures of hard disks with a size-dependent non-additivity parameter $\Delta$ given by:
\begin{equation}
    \Delta = \frac{\sigma_{LS}}{(\sigma_L + \sigma_S)/2} - 1 = \frac{2\sqrt{q}}{1 + q} -1.
\end{equation} 
For each size ratio, we use the data from Ref. \onlinecite{fayen2020infinite} for the best packed candidate structures, which were obtained from systematic sampling of unit cells containing up to 12 particles using Monte Carlo simulations with a variable box shape \cite{filion2009prediction}. The infinite pressure phase diagram is then constructed from these structures by common-tangent construction \cite{fayen2020infinite}. We show the result in Fig. \ref{fig:infinitepressure} for size ratios $q$ between $0.25$ and $0.75$. Note that for large size ratios, $\Delta$ tends to 0 and the system becomes almost additive. In the additive case, it is proven that there exist no denser structures than a coexistence of Hex$_L$ and Hex$_S$ for $q \gtrsim 0.74$ \cite{blind_gerd_uber_1969}. Therefore we expect no additional binary crystal phases to appear beyond size ratio $q>0.75$, and cut the infinite pressure phase diagram at this point.
In addition to the trivial monodisperse hexagonal crystal phases of the large or small particles (\HexL{} and \HexS{}, respectively), we observe a wide variety of binary phases. Since any pure crystal phase can only occur at a single composition $x_S$, the densest-packed state at most points in the phase diagram (white regions) is a coexistence between two crystal phases at different compositions: the ones appearing directly above and below the chosen state point. 

For each binary phase in Figure \ref{fig:infinitepressure} we also depict the repeating unit that can be used to construct the crystal phase, which we call a tile.  Unlike a unit cell, tiles can appear in the full crystal structure in multiple orientations.  Interestingly, for certain coexistence regions, the two coexisting phases consist of tiles that can mix.  One realization of this occurs at low size ratios ($q \simeq 0.3$) where the T1 and \HexL{} phases consist of identical triangles of large particles, but decorated differently by small particles.  In the region where these two phases mix, they can form a lattice gas where tiles of T1 and \HexL{} are randomly mixed.  Another, much more interesting situation occurs when two tiles of different shapes can mix.  This occurs in the three hashed regions in Fig. \ref{fig:infinitepressure}. For example, at size ratios just below $q=0.5$, the square tile of the S1 phase has the same edge length as the  triangular tile of the \HexL{} phase, allowing them to mix and form a space-filling square-triangle tiling \cite{likos1993complex, fayen2020infinite}, illustrated in the bottom right of Fig. \ref{fig:infinitepressure}. As this mixing increases the entropy without lowering the packing fraction, the expected phase at infinite pressure here is a random tiling of squares and triangles, which at an ideal composition $x_S = (3-\sqrt{3})/4 \simeq 0.317$ is known to have 12-fold symmetry on average \cite{oxborrow1993random, imperor2021square}. Two closely related tilings, also illustrated in Fig. \ref{fig:infinitepressure} are found at lower size ratios.  As a result, one intriguing prediction from Fig. \ref{fig:infinitepressure} is the possibility of a 12-fold quasicrystal self-assembling from simple binary mixtures of colloidal spheres on a substrate.

\section{Finite pressure self-assembly}

In practice, the infinite-pressure phase behavior is not a reliable indication for the phases one might find in a real self-assembly experiment. Self-assembly in a colloidal system takes place at finite pressure, where contributions from the vibrational entropy to the free energy of different crystal phases can fundamentally change the phase behavior. Moreover, dynamical arrest or competition with other candidate phases can prevent the reliable formation of a crystal even if it is thermodynamically stable.

Hence, for a more realistic look at the self-assembly, we perform computer simulations  at finite pressure at an extensive grid of state points spanning size ratios $0.25 \leq q \leq 0.75$, compositions $0.05 \leq x_S \leq 0.95$, and packing fractions $0.7 \leq \eta \leq 1.0$.
In particular, we run event-driven molecular dynamics (EDMD) simulations \cite{rapaport2004art} in the canonical ensemble, i.e. at constant number of particles $N$, volume $V$, and temperature $T$. The simulation algorithm is a variant of the methods described in Ref. \cite{smallenburg2022efficient}. Initial configurations consisting of $N_L + N_S = 2000$ particles at different packing fractions were obtained by starting in a dilute state at the desired composition, and then performing an EDMD simulation in which the particle diameters grow until the desired packing fraction is reached. Subsequently, each system is allowed to evolve at constant volume for at least $10^6 \tau_\mathrm{MD}$, where $\tau_\mathrm{MD} = \sqrt{m\sigma_A^2/k_BT}$ is the time unit of our simulation, with $m$ the mass of a particle, $k_B$ Boltzmann's constant, and $T$ the temperature. For each run, we examine the final configuration as well as the diffraction pattern obtained from the Fourier transform of the large-particle coordinates in order to detect crystallization.

\begin{figure*}
    \centering
    \includegraphics[width=\textwidth]{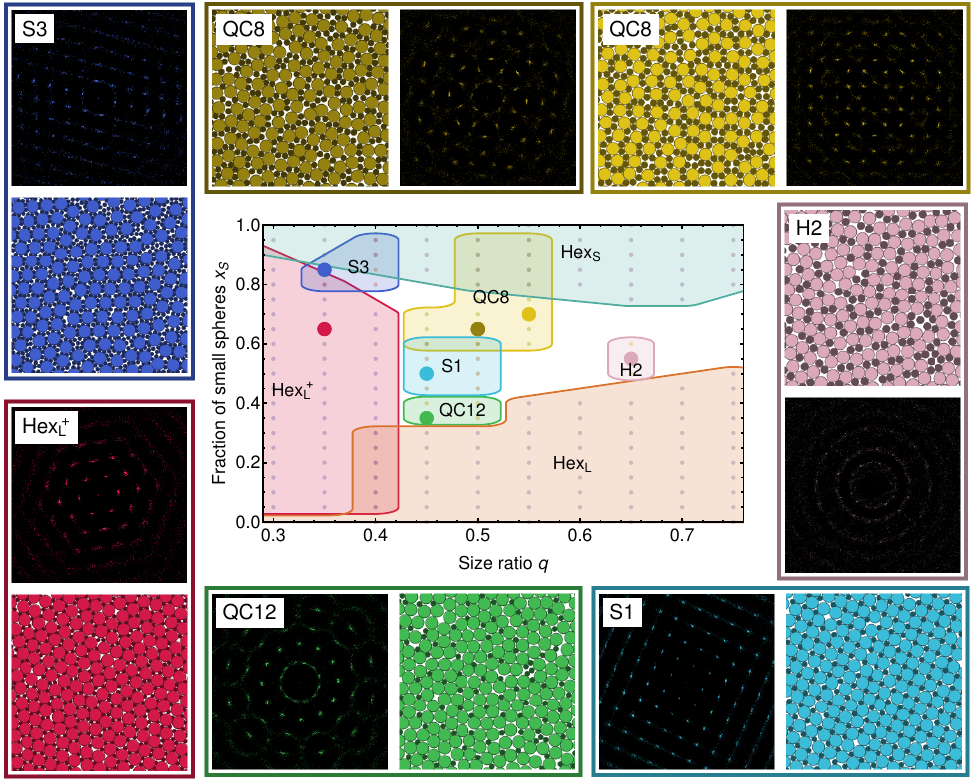}
    \caption{Self-assembly diagram for binary mixtures of spheres on a flat plane, as a function of the size ratio $q$ and fraction of small particles $x_S$. For each combination of $q$ and $x_S$, we perform simulations at a range of different packing fractions, and report the observed phases. A colored point in the phase diagram indicates the self-assembly of the corresponding phase. At state points where no point is shown, no crystallization was detected. For each binary crystal phase, we include a typical snapshot and the scattering pattern that results from a Fourier transform of the positions of the large spheres. For the QC8 phase, we include two snapshots: one containing a large concentration of S1 squares (top middle) and one containing a large concentration of S2 squares (top right). \HexL{} and \HexS{} are hexagonal crystals consisting of only large or small spheres, respectively, and are not depicted.}
    \label{fig:selfassembly}
\end{figure*}

Our results are summarized in Fig. \ref{fig:selfassembly}. The central diagram reports for each investigated combination of $q$ and $x_S$ what ordered phases were observed. We consider a crystal to have self-assembled for a given combination of $q$ and $x_S$ when we find significant clusters of the crystal in the simulation box for at least one packing fraction. At state points without an indicated crystal phase, no crystallization was observed at any of the investigated packing fractions.  For the quasicrystals, local crystalline order is typically hard to see by eye, and we instead rely on the symmetry of the scattering pattern for our classification. Note that in the Supplemental Information (SI), we include a full catalogue of all final configurations and their diffraction patterns.

\begin{figure*}
    \centering
    \includegraphics[width = 0.9\linewidth]{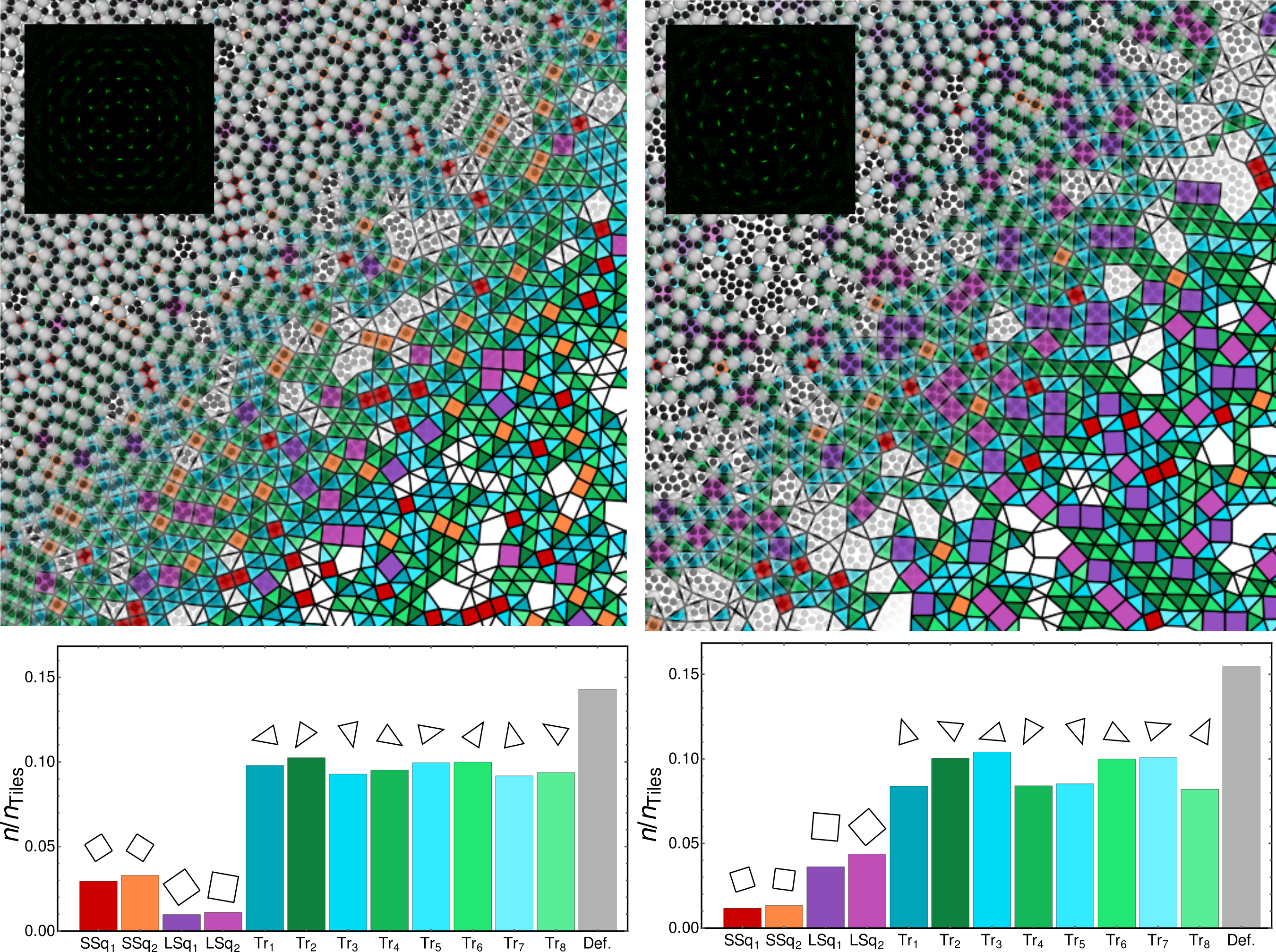}
    \caption{Self-assembled octagonal random tiling quasicrystals in mixtures of $10^4$ spheres on a flat plane, at state points corresponding to the QC8 phase with different concentrations of large and small squares. The underlying tilings are highlighted and colored according to shape and orientation. The insets show the diffraction patterns, signaling the global 8-fold symmetry. Tile distributions show that all tile orientations of the same shape appear with similar frequencies. The bar labeled ``Def.'' denotes all defects. The state points are: (Left) $q = 0.5$, $x_S = 0.675$ and $\eta = 0.86$. (Right) $q = 0.55$, $x_S = 0.715$ and $\eta = 0.84$.}
    \label{fig:tiling_analysis}
\end{figure*}

Our simulations show that a number of the best-packed phases we predicted in Fig. \ref{fig:infinitepressure}  indeed spontaneously self-assemble. Naturally, this includes the trivial hexagonal crystals of the large and small spheres (\HexL{} and \HexS{}) that can be found at compositions close to $x_S = 0$ and $1$, respectively. Additionally, we observe large-scale crystallization into the S1 and S3 phases close to the regions expected from Fig. \ref{fig:infinitepressure}. We also observe the more complex H2 phase, albeit only in finite clusters -- a closer inspection of the systems where these form show a very low overall mobility of the system, suggesting that crystallization of this phase is hindered by slow dynamics. For sufficiently low $q$, the system nearly always forms a hexagonal lattice of large spheres, with the small spheres interspersed between them (labeled \HexLplus{}). Depending on the composition, this may look similar to the T1 phase (as depicted in the sample snapshot in Fig. \ref{fig:selfassembly}), but the number of small spheres per triangular cavity in the lattice of large spheres appears to continuously depend on the composition $x_S$ (see SI). For $x_S<2/3$, this simply means that a random selection of the triangular holes are empty, resulting in a lattice gas or interstitial solid solution \cite{fayen2020infinite, likos1993complex}. For larger $x_S$, progressively more small particles are included between the large spheres, but we observe no clear structural transition between these regimes. Hence, we choose to collectively indicate this region as \HexLplus{}.

Most intriguingly, in addition to these periodic phases, we also observe the self-assembly of two distinct quasicrystals, both at size ratios between $q=0.45$ and $q=0.55$. The dodecagonal quasicrystal (QC12) that appears at low fractions of small spheres is indeed the square-triangle tiling \cite{oxborrow1993random,imperor2021square} expected from the infinite-pressure diagram. It is made of regular squares and triangles (S1 and \HexL{} tiles). This quasicrystal is analogous to a number of quasicrystals observed in soft matter systems, including patchy particles with five attractive patches \cite{van2012formation}, hard disks with a square-shoulder repulsion \cite{dotera2014mosaic, pattabhiraman2015stability}, binary mixtures of nanoparticles \cite{talapin2009quasicrystalline}, block copolymers \cite{gillard2016dodecagonal, duan2018stability}, and soft repulsive colloids \cite{zu2017forming}. Additionally, various 3D systems have been shown to form quasicrystals consisting of layers of a square-triangle tiling \cite{haji2009disordered, ye2017quasicrystalline, haji2011degenerate}.

The second quasicrystal (QC8) has octagonal symmetry, and consists of a mixture of three tiles: the isosceles triangles that appear in the H1 phase, the squares from the S1 lattices, and the larger squares from the S2 lattice. In order to examine this structure in more detail, we perform additional simulations of $N=10000$ particles in the regime where this phase was found to self-assemble, and analyze the tilings in the final configurations. To identify the underlying tiling in simulation snapshots, we first draw bonds between all large particles that are closer than a cutoff distance $r_c = 1.7 \sigma_{LL}$. After removal of the crossing bonds, which typically occur in S1 unit cells, we compute the bond length and orientation distributions. At state points where the octagonal quasicrystal self-assembles, the bond length distribution is clearly bimodal and a cutoff can be defined to discriminate between short and long bonds. Similarly, the bond orientation distribution exhibits 16 sharp peaks that alternatively correlate very strongly with long and short bonds. We provide examples of these distributions in the SI. Tiles are then reconstructed by  cycling through  bonds, and, subsequently, sorted by shape and orientation.

In Figure \ref{fig:tiling_analysis}, we show portions of the final state of two simulated mixtures of $10^4$ particles, at different state points. The left one is dominated by small squares, while the second one, which contains more small particles, predominantly contains large squares. Both systems possess global octagonal symmetry as indicated by the diffraction patterns.
The analysis of the tile orientations shows that for a given shape -- small squares, large squares and isosceles triangles -- all possible orientations appear roughly with the same frequency, which is a common feature of random tiling quasicrystals \cite{imperor2021square}.

As illustrated in Figure \ref{fig:tiling_analysis}, the relative concentrations of the different tiles in the QC8 phase vary drastically as a function of the fraction of small spheres in the system. Since the S2 squares contain 4 small particles each, while the S1 squares only contain a single small sphere, higher compositions $x_S$ favor a larger concentration of S2 squares. For high $x_S$, the QC8 tiling consists almost purely of large S2 squares and H1 triangles, with the triangles joined in pairs that form a thin rhombus. In this limit, the tiling can be seen as a mixture of just two types of tiles -- square and rhombic -- that are identical to the tiles that form  e.g. the Ammann-Beenker \cite{ammann1992aperiodic, beenker1982algebraic} and Watanabe-Ito-Soma \cite{watanabe1987nonperiodic} octagonal aperiodic tilings. The same tiling -- with different decorations of the tiles with particles -- was previously observed in simulations of soft colloids \cite{zu2017forming}, particles with an oscillating interaction potential \cite{damasceno2017non, malescio2021self}, and patchy particles \cite{gemeinhardt2019stabilizing}. However, to our knowledge, no octagonal quasicrystals have been observed to spontaneously self-assemble in soft-matter experiments.

%Note that in our system, a perfect Ammann-Beenker or Watanabe-Ito-Soma tiling would correspond to a composition $x_S = (4\sqrt{2}+2)/(5\sqrt{2}+3) \simeq  0.7603$.

On the other hand, at low $x_S$ the quasicrystal approaches a tiling of only H1 triangles and small S1 squares. This can be seen as a separate two-tile random-tiling quasicrystal which, to our knowledge, has not previously been observed in soft matter systems. Interestingly, however, a closely related tiling, where the isosceles triangles are slightly deformed, was recently conjectured to be the densest-packed structure for a ternary mixture of hard disks \cite{fernique2021compact}.

\begin{figure}
    \centering
    \includegraphics[width=\linewidth]{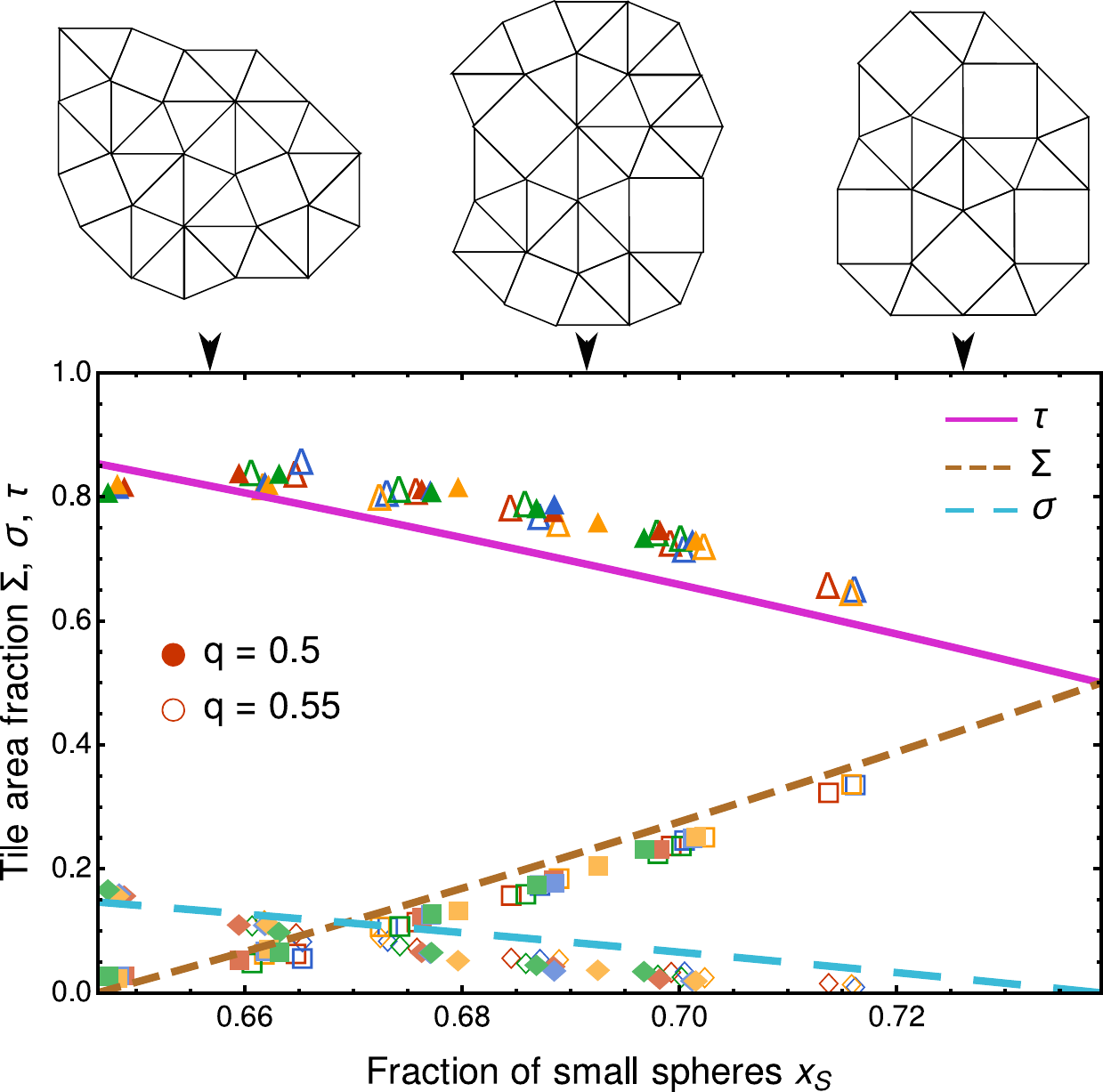}
    \caption{
    Area fractions of the three different tiles in the QC8 tiling, $\Sigma$, $\sigma$, and $\tau$, corresponding to the large squares, small squares, and triangles, respectively. The lines indicate the theoretical prediction on the assumption of a maximally symmetric and globally uniform eight-fold tiling with no defects. Points correspond to simulation results at size ratios $q=0.5$ (full symbols) and $q=0.55$ (open symbols) as indicated. Different colors of points correspond to different packing fractions, with $0.855 \leq \eta \leq 0.87$ for $q = 0.5$ and $0.835 \leq \eta \leq 0.85$ for $q = 0.55$. For the simulation data, we only consider the area covered by non-defect tiles when calculating the composition $x_S$ and the tile area fractions.
    At the top, three patches illustrate the evolution of the tilings with the composition. From left to right: primarily small squares, mixture of small and large squares and primarily large squares.
    }
    \label{fig:QC8composition}
\end{figure}

It is interesting to consider under what conditions the QC8 tiling observed here can exhibit true 8-fold symmetry. Counting different orientations, this tiling consists of 12 different tiles: 2 orientations of large squares, 2 orientations of  small squares, and 8 differently oriented triangles. The lattice of vertices can be seen as a projection of a four-dimensional hyperlattice, and as a result, any particular tiling can be ``lifted'' to a set of points in four dimensions that forms a subset of this hyperlattice \cite{baake_grimm_2013, baake2016guide}. As outlined in Ref. \onlinecite{imperor2021square}, this lifting procedure can aid in determining the constraints  on the relative concentrations of different tiles. As described in the SI, for a QC8 with octagonal symmetry we find the following constraint for the partial area fractions $\Sigma$, $\sigma$, and $\tau$, associated with the large S2 squares, small S1 squares, and the triangles that make up H1, respectively:
\begin{equation}
   \Sigma + (3 + 2\sqrt{2})\sigma - \tau = 0.
\end{equation}
Additionally, these area fractions must trivially satisfy $\Sigma+\sigma+\tau = 1$.
Since we know the composition of each tile in our binary mixture, it is straightforward to rewrite these constraints in terms of the fraction of small particles $x_S$, yielding:
\begin{eqnarray}
    \Sigma &=&\frac{2 \left(4+3 \sqrt{2}\right) x_S - 4 \sqrt{2} - 5}{6 - 4 x_S}\\
    \sigma &=& \frac{-\left(4+\sqrt{2}\right) x_S + 4}{6-4 x_S}\\
    \tau &=&\frac{-\left(8+5 \sqrt{2}\right) x_S+4 \sqrt{2}+7}{6-4 x_S}
\end{eqnarray}
In Fig. \ref{fig:QC8composition}, we draw this prediction together with the measured tile concentrations in our self-assembled configurations of 10000 particles. Note that in the analysis of the simulation data, we consider only the portion of the system covered by the three valid types of tiles and omit all defects. We find that the observed tile concentrations are essentially independent of size ratio and packing fractions within the investigated regime. Considering the fact that the analyzed configurations were the result of spontaneous self-assembly, and hence contain significant amounts of defects, the agreement is excellent, demonstrating that the system indeed favors tile compositions that correspond to an eight-fold quasicrystalline symmetry. 

\subsection{Geometrical constraints}

\begin{figure}
    \centering
    \includegraphics[width = \linewidth]{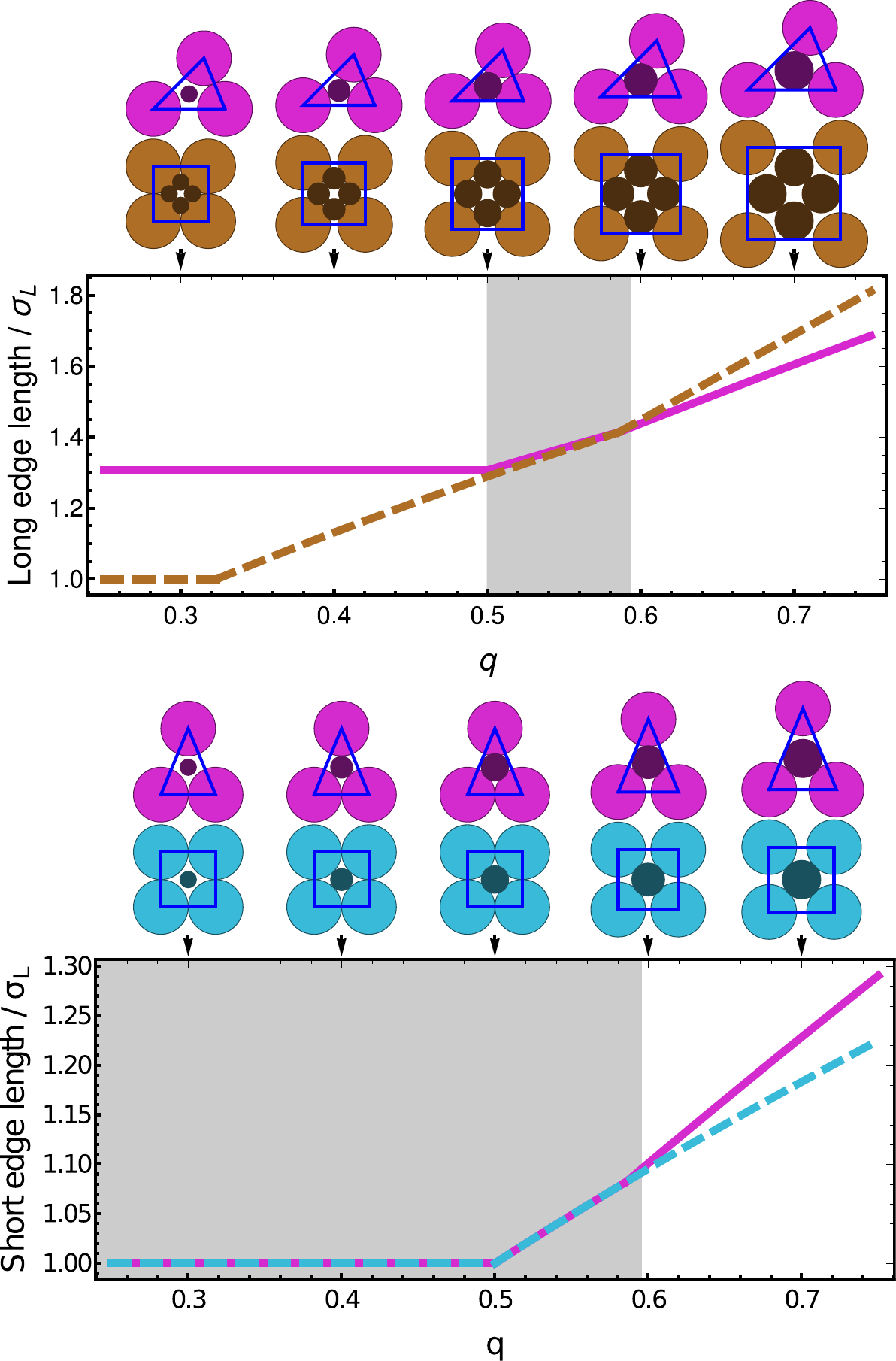}
    \caption{Evolution of the possible long (top) and short (bottom) edge lengths as a function of size ratio. Matching regions are highlighted with a dark background. For size ratios between 0.5 and 0.6, long edges of H1 and S2 on one hand, and short edges of H1 and S1 on the other hand match, thus allowing for the tiles that comprise the octagonal tiling to mix. Self-assembly of QC8 is indeed observed for these values of the size ratio.}
    \label{fig:QCs}
\end{figure}

An intriguing question remains --  is there a way to understand why these octagonal quasicrystals appear in this highly simple system? As stated, the three tiles that comprise the tiling are the small S1 square, the large S2 square, and the H1 triangle. In order to form the observed tilings, these shapes must have compatible edge lengths on their shared edges. In particular, the shared edges in the observed tilings are between the S2 square and the long edge of the H1 triangle, and the S1 square and the short edge of the H1 triangle. As shown in Figure \ref{fig:QCs}, the long edge of the H1 triangle matches up almost exactly with the edge of an S2 square for size ratios between 0.5 and 0.6, in the region where we observe the self-assembly of this phase. Similarly, the short edge of the H1 triangle and the S1 square match exactly for size ratios below $q = 2-\sqrt{2} \simeq 0.59$. The fact that a QC8 with mainly small squares is not observed at size ratios below $q=0.45$ can be understood from the observation that for $q<1/2$, the small particle in both the H1 triangle and the S1 square are no longer touching their neighbors. Hence, as $q$ decreases, the packing fraction of these tiles decreases rapidly, causing them to become unfavorable in comparison to differently shaped tiles which pack better.

\section{Conclusion}

In conclusion, we have explored the self-assembled phases that appear in binary systems of hard spheres on a flat plane. In addition to a variety of periodic crystals, we found that this very simple system is capable of forming two different quasicrystal structures: one dodecagonal and one octagonal. The octagonal one is particularly intriguing, as it consists of three distinct tiles, whose relative concentration can be continuously tuned by manipulating the prevalence of small spheres. Both observed quasicrystals self-assemble rapidly and reliably over a significant region of parameter space and their stability can be readily understood from geometrical arguments. In contrast to nearly all other interaction potentials that have been shown to form 2D quasicrystals, this hard sphere model has been shown to be quantitatively reproducible in the lab \cite{thorneywork2014communication, thorneywork2017two}. Hence, this system is arguably the ideal model for exploring soft matter quasicrystal self-assembly.

\section*{Acknowledgements}
We thank Thomas Fernique,  Jean-Fran\c{c}ois Sadoc, Pavel Kalouguine, Alptu\u{g} Ulug\"ol, and Anuradha Jagannathan for many useful discussions. EF, GF, and FS acknowledge funding from the  Agence Nationale de la Recherche (ANR), grant ANR-18-CE09-0025. LF acknowledges funding from the Netherlands  Organisation  for  Scientific  Research (NWO) for a Vidi grant (Grant No. VI.VIDI.192.102)

\section*{References}

\bibliography{refs}

\end{document}